\documentclass[11pt]{elsart}

\include{latexsym}
\include{epsf}
\def\DoubleR{{\rm\bf R}}

\begin{document}

\begin{frontmatter}

\title{Optimal embedding parameters: A modeling paradigm}
\author{Michael Small\thanksref{correspond}}
\author{C.K. Tse}
\address{Department of Electronic and Information Engineering\\
The Hong Kong Polytechnic University\\
Hung Hom, Kowloon, Hong Kong}
\thanks[correspond]{Corresponding author. Tel: +852 2766 4744,
Fax: +852 2362 8439, email: {\tt ensmall@polyu.edu.hk}.}

\date{\today} 
\maketitle

\begin{abstract}
Reconstruction of a dynamical system from a time series requires the
selection of two parameters, the embedding dimension $d_e$ and the
embedding lag $\tau$. Many competing criteria to select these parameters
exist, and all are heuristic. Within the context of modeling the
evolution operator of the underlying dynamical system, we show that one
only need be concerned with the product $d_e\tau$. We introduce an
information theoretic criteria for the optimal selection of the
embedding window $d_w=d_e\tau$. For infinitely long time series this
method is equivalent to selecting the embedding lag that minimises the
nonlinear model prediction error. For short and noisy time series we
find that the results of this new algorithm are data dependent and
superior to estimation of embedding parameters with the standard
techniques.
\end{abstract}

\end{frontmatter}

\section{Reconstruction}

The celebrated theorem of Takens \cite{fT81} guarantees that, for a
sufficiently long time series of scalar observations of an
$n$-dimensional dynamical system with a $C^2$ measurement function, one
may recreate the underlying dynamics (up to homeomorphism) with a time delay embedding\footnote{Takens' theorem has many extensions and is
described in various forms by several contemporary authors. We do not
intend to dwell on the evolution of these fundamental results
here.}. Unfortunately the theorem is silent on exactly how to proceed
when the data is limited and contaminated by noise. In practice, time
delay embedding is routinely employed as a first step in the analysis of
experimentally observed nonlinear dynamical systems (see
\cite{hA96b,hK97}). Typically, one identifies some characteristic
embedding lag $\tau$ (usually related to the sampling rate and time
scale of the time series under consideration) and utilises $d_e$ lagged
version of the scalar observable for sufficiently large $d_e$. In
general, $\tau$ is determined by identifying linear or nonlinear
temporal correlations in the data and one will progressively increase
$d_e$ until the results obtained are self consistent. 

In this paper we consider the problem of reconstructing the underlying
dynamics from a finite scalar time series in the presence of noise. We
recognise that in general the quality of the reconstruction will depend
on the length of the time series and the amount of noise present in the
system. Employing the minimum description length model selection
criteria we show that the optimal model of the dynamics does not depend
on the choice of the embedding lag, only on the maximum lag ($d_e\tau$
in the above scheme). We call that maximum embedding lag $d_w:=d_e\tau$, the
embedding window, and show that for long noise-free time series the
optimal $d_w$ minimises the one-step model prediction error. For short
or noisy data, the optimal value of $d_w$ is data dependent. To estimate
the one-step model prediction error and $d_w$ we apply a generic local
constant modeling scheme to several computational examples. We show
that this method proves to be consistent and robust, and the results that we
obtain capture the salient features of the underlying dynamics. Finally,
we also find that in general there is no single
characteristic time lag $\tau$. Generically, the optimal
reconstruction may be obtained by considering the lag vector
\begin{eqnarray}\label{varemb}
(\tau_1,\tau_2,\ldots,\tau_k)
\end{eqnarray} where $0<\tau_i<\tau_{i+1}\leq d_w$
\footnote{This is the so called ``variable embedding'' described in
\cite{kJ98} and elsewhere.}. 

The textbooks \cite{hA96b,hK97} contain copious detail on the
estimation of $d_e$ and $\tau$. We briefly review only the most relevant
developments here.

Often, the primary aim of time delay embedding is the estimation of
dynamic invariants. In these instances, one may estimate $\tau$ with a
variety of heuristic techniques: usually autocorrelation, pseudo-period or
mutual information. One then computes the dynamic invariant for increasing
values of $d_e$ until some sort of plateau onset occurs (see \cite{hK98}
and the references therein). For estimation of correlation
dimension, $d_c$, it has been shown that $d_e>d_c$ is sufficient
\cite{grebogi:dim-est}. However, for reconstruction of the underlying
dynamics this is not the case. Alternatively, the method of false
nearest neighbours \cite{mK92} and its various extensions apply a
topological reasoning: one increases $d_e$ until the geometry of the
time series does not change.

We note that several authors have speculated on whether the individual
parameters $d_e$ and $\tau$, or only their product $d_e\tau$, is
significant. For example, Lai and Lerner \cite{yL98} provide an overview
of selection of embedding parameters to estimate dynamic invariants (in
their case, correlation dimension).  They impose some fairly generous
constraints on the correlation integral and use these to estimate the
optimal value of $d_e$ and $\tau$. Their numerical results from long
clean data imply that correct selection of $\tau$ is crucial, selection
of $d_w$ (and therefore $d_e$) is not. Conversely, utilising the BDS
statistic \cite{wB91}, Kim and co-workers \cite{hK98} concluded that the
crucial parameter for estimating correlation dimension is $d_w$.

Unlike these previous methods, the question we consider is: ``What is
the optimal choice of embedding parameters to reconstruct the underlying
dynamic evolution from a time series?'' In answering this question we
conclude that only the embedding window $d_w$ is significant, selection
of optimal embedding lags is, essentially, a modeling problem
\cite{kJ98}. Clearly, successful reconstruction of the underlying
dynamics will depend on ones' ability to identify any underlying
periodicity (and therefore $\tau$). The results of this paper shows that
it is possible to estimate the optimal value of $d_w$, and subsequently
use this optimal value to derive a suitable embedding lag
$\tau$. However, as previous authors have observed in many examples,
estimation of $\tau$ for nonlinear systems is model dependent
\cite{kJ98} (and may even be {\em state} dependent).

In the following section we introduce our main result and the rational
for the calculations that follow. Section \ref{examples} demonstrates
the application of this method to several test systems, and section
\ref{applications} studies the problem of modelling several experimental
time series. In section \ref{conclusions} we conclude.

\section{A modeling paradigm}
\label{paradigm}

Let $\phi:\mathcal{M}\longrightarrow\mathcal{M}$ be the evolution operator of a
dynamical system, and $h:\mathcal{M}\longrightarrow\DoubleR$ a $C^2$
differentiable observation function. Through some experiment we obtain
the time series $\{h(X_1),h(X_2),\ldots,h(X_N)\}$. Denote $x_i\equiv
h(X_i)$. Takens' theorem
\cite{fT81} states that for some $m>0$ the mapping $g$
\begin{eqnarray}
\label{embed0}
x_i  & \stackrel{g}{\longmapsto} &
(x_i,x_{i-1},x_{i-2},\ldots,x_{i-m-1})
\end{eqnarray}
is such that the evolution of $g(x_i)=(x_i,x_{i-1},\ldots,x_{i-m-1})$\footnote{In writing $g(x_i)=(x_i,x_{i-1},\ldots,x_{i-m-1})$ we take
a slight liberty with the notation, but the meaning remains clear.} is
homeomorphic to $\phi$  

We will generalise the embedding map (\ref{embed0}) and consider $\hat{g}$ as
\begin{eqnarray}
\label{embed}
x_i \stackrel{\hat{g}}{\longmapsto} (a_1x_i,a_2x_{i-1},a_3x_{i-2},\ldots,a_dx_{i-d-1}).
\end{eqnarray}
The objective of a successful embedding is to find
$a=[a_1,a_2,\ldots,a_d]$ where $a_i\in\{0,1\}$. Note that
$\hat{g}(x_i)$ is simply the subspace projection of $g(x_i)$ onto $a$,  
\[
\hat{g}(x_i)=\mbox{Proj}_ag(x_i).
\]
The embedding is completely defined by $a\in\{0,1\}^d$ and we wish to
make the best choice of $a$ and $d$, which we write $(a,d)$. Note that,
in general one could consider $a\in\DoubleR^{d_e\tau}$. We restrict
ourselves to $\{0,1\}^d$ as the more general case is concerned with the
optimal model of the dynamics rather than the necessary information. For a
uniform embedding with embedding parameters $d_e$ and $\tau$ we have that 
$a\in\{0,1\}^d$ and $(a)_i\neq0$ if and only if $\tau$ divides $i$.

Let $z_i=\hat{g}(x_i)\in\DoubleR^d$ and let
\begin{eqnarray}
\label{func}
f(z)&=&\sum_{i=1}^m\lambda_i\theta(z;w_i)
\end{eqnarray}
where $\theta$ is some basis and $\lambda_i\in\DoubleR$ and
$w_i\in\DoubleR^k$ are linear and nonlinear model parameters. The
selection of this particular model architecture is arbitrary, but does
not alter the results. We assume
that there exists some algorithm to select $\mathcal{P}=(m,\lambda_1,\lambda_2,\ldots,\lambda_m,w_1,w_2,\ldots,w_k)$ such
that $e_i=f(z_{i-1})-z_i\sim N(0,\sigma)$ (or at the very least, $\sum
(f(z_{i-1})-z_i)^2=\sigma^2$ is 
minimised). We do not consider the model selection problem here, rather
we seek to find out what is the best choice of $(a,d)$. Our own model
selection work is summarised in \cite{mdlnn}.

The most obvious approach to this problem is to look for the maximum
likelihood solution:
\[\max_{(a,d)}\max_\mathcal{P}P(x|x_0,a,d,\mathcal{P})\]
where $x$ is the vector of all the time series observations and $x_0\in\DoubleR^d$ is a vector
of model initial conditions. Unfortunately this leads to the redundant
solution $d=N$. To solve this problem one could either resort to
Bayesian regularisation \cite{dM92} or the minimum description length model
selection criteria \cite{jR89}. We choose the later approach.

The description length of a time series is the length of the shortest
(most compact) 
description of that time series. The description length of a time series
with respect to a given model is the length of the description of that
model, the initial conditions of that model and the model prediction
error. We intend to optimise the description length of the observed time
series $\{x_i\}_{i=1}^N=x$ with respect to $(a,d)$. At this point we
make the fairly cavalier assumption that for a
given $(a,d)$ one can obtain the optimal model $\mathcal{P}$. We will
address this assumption in more detail later in this section. 

The description length
of the data $DL(x)$ is given by
\begin{eqnarray}
\label{dl1}
DL(x) & = & DL(x|x_0,a,d,\mathcal{P}) + DL(x_0) + DL(a,d) + DL(\mathcal{P})
\end{eqnarray}
where $x_0=(x_1,x_2,\ldots,x_d)$ are the model initial
conditions. Notice that the description
length of the model prediction errors $DL(x|x_0,a,d,\mathcal{P})$, is  equal to the
negative log likelihood of the errors under the assumed
distributed. Similarly $x_0$ is a sequence of $d$ real numbers which for
small $d$ we approximate by $d$ realisations of a random
variable. Therefore $DL(x_0)$ can also be computed as a negative
log-likelihood of some probability distribution. If we assume that $x$
and $x_0$ are approximated by Gaussian random variables with variance
$\sigma^2$ and $\sigma_D^2$ respectively, then (\ref{dl1})
becomes
\begin{eqnarray}
\label{dl2}
DL(x) & \approx & -\ln{P(x|N(0,\sigma^2))} \\
\nonumber & & \mbox{  } - \ln{P(x_0|N(0,\sigma_X^2))} + d + DL(d) + DL(\mathcal{P}).
\end{eqnarray}
Since $a$ is a sequence of $d$ independent zeros or ones
$DL(a)=d$, furthermore the description length of an integer $d$ is given
by
$DL(d)=\lceil\log(d)\rceil+\lceil\log\lceil\log(d)\rceil\rceil+\ldots$
where the last term in this expansion is $0$\cite{jR89}. Compared to the term
$d$, $DL(d)$ is very slowly varying and has little effect on the
results. The final term $DL(\mathcal{P})$ is the description length of
the optimal model for the given $(a,d)$. 

Substituting for the probability distributions
$P(x|N(0,\sigma^2))$ and $P(x_0|N(0,\sigma_X^2))$ and estimating
$\sigma^2$ and $\sigma_X^2$ directly from the data, one finally obtains
\begin{eqnarray}
\label{dl3}
DL(x) & \approx & 
\frac{d}{2}\left(1 + \ln{2\pi\sigma_X^2}\right)
+\frac{N-d}{2}\left(1 + \ln{2\pi\sigma^2}\right)\\
\nonumber && \mbox{  }
+ DL(\overline{x})+d+DL(d)+DL(\mathcal{P})\\
\nonumber && \mbox{ }\\
\label{dl4}
& \approx &
\frac{d}{2}\ln{\left[\frac{1}{d}\sum_{i=1}^d(x_i-\overline{x})^2\right]}
+\frac{N-d}{2}\ln{\left[\frac{1}{N-d}\sum_{i=d+1}^Ne_i^2\right]}\\
\nonumber & & \mbox{  }
+\frac{N}{2}\left(1+\ln{2\pi}\right)+d+ DL(d)+DL(\overline{x})+DL(\mathcal{P}).
\end{eqnarray}
In this form, equation (\ref{dl4}) provides the first suggestion of what
the optimal embedding strategy should be. We see that $a$ does not
feature in this calculation. Hence, if we adopt the modeling paradigm
suggested here, the embedding lag (or more generally the embedding
strategy) is not crucial: one should only be concerned with the
maximum embedding dimension $d$. Of course, this does not mean that to
reconstruct the dynamics the
embedding lag is  unimportant. When one applies numerical modeling to
reconstruct the dynamics, embedding strategies are of very great
significance, however selection of the optimal embedding co-ordinates
(or rather those that are most significant in predicting the dynamics)
is inherently part of the modeling process \cite{kJ98}. Furthermore, the
modelling algorithm should be allowed to choose from all possible
embedding lags within the embedding window. Indeed, one often finds that
the ``optimal'' embedding strategy is not fixed within a single model
\cite{kJ98}. This result shows that it is preferable to identify the
embedding window $d_w$ and let the model building process determine
which of the $d_w$ co-ordinates are most useful.  

The description length of the mean of the data $DL(\overline{x})$ is a
fixed constant and we drop it from the calculation.  Optimising
(\ref{dl4}) over all $(a,d)$ requires selection of the optimal model for
a given $(a,d)$ and computation of the model prediction error of that
model. For a given model,  $DL(\mathcal{P})$ can be calculated precisely
\cite{kJ95a}. However, selection of the optimal model is a more
difficult problem. 

Instead, we restrict our attention to a particular {\em class} of model,
and choose the optimal model from that class. To simplify the
computation of $(\ref{dl4})$ we restrict our attention to the class of
local constant models on the attractor. We have two good reasons for
choosing this particular class. Firstly, because the models are
simple, estimates of the error as a function of $(a,d)$ are relatively well
behaved. Secondly, these models rely on no additional parameters and therefore
$DL(\mathcal{P})=0$, simplifying our calculation
considerably\footnote{Alternatively, one could argue that the data are
the parameters, in either case the description length of the model is
constant.}.

In trials, we tested many alternative model classes. We found radial
basis functions \cite{kJ95a} and neural networks \cite{mdlnn} to be
excessively nonlinear and difficult to optimise for the purpose of
determining embedding windows. Complex local modeling regimes such as
triangulation and tessellation \cite{aM91} or parameter dependent local
linear schemes \cite{rH99} we found to be overly sensitive to small
changes in the data. In comparison the local constant scheme we employ
here appears remarkably robust.

As local constant models have no explicit parameters (other than the
embedding strategy $(a,d)$), $DL(\mathcal{P})=0$. Therefore, for a given $(a,d)$
computation of $(\ref{dl4})$ only requires estimation of $\sum e_i^2$.
We employ an in-sample local constant prediction strategy. Let $z_s$ be
the nearest neighbour to $z_t$ (excluding $z_t$), then 
\begin{eqnarray}
\label{model}
x_{t+1} & = & x_{s+1}
\end{eqnarray}
and therefore $e_{t+1}=x_{t+1}-x_{s+1}$. In other words, for each point
in the time series we determine the prediction error based on the
difference between the successor to that point and the successor to its
nearest neighbour\footnote{This is a technique sometimes referred to as
``drop-one-out'' interpolation.}. Since this is a form of interpolation
rather than extrapolation, this strategy does not provide a {\em
predictive} model, likewise (as with all local techniques) it does not
describe the underlying dynamics. However, the strength of this
particular approach is that it is simple and it provides a realistic
estimate of the size of the optimal model's prediction error as a
function of $(a,d)$.

The proposed algorithm may be summarised as follows. We seek to minimise
(\ref{dl4}) over $d$. To achieve this we need to estimate the model
prediction error as a function of $d$. Hence, for increasing values of
$d$ we employ the local constant ``modelling'' scheme suggested by
(\ref{model}) to compute the model prediction error and substitute this
into (\ref{dl4}). The optimal embedding window $d_w$ is the value of
$d$ that minimises (\ref{dl4}).

\section{Examples}
\label{examples}

In this section we describe the application of the above method to
several numerical time series. First, we examine the
performance of the algorithm and importance of the choice of modelling
algorithm (\ref{model}). 

\begin{figure}[t]
\[\epsfxsize 125mm \epsfbox{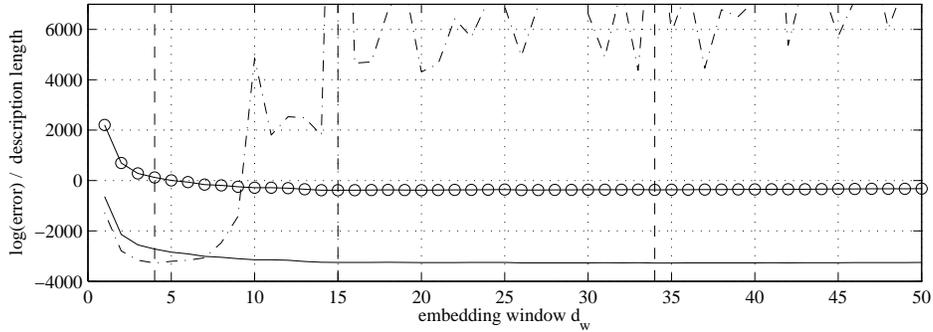}\]
\caption{{\bf Computation of description length as a function of
embedding window for R\"ossler time series data.} The solid line and
dot-dashed line are proportional to the logarithm of the sum of the
squares of the
model prediction error using a local constant and local linear method
respectively. The local constant model utilised is described in section
\ref{paradigm}, the local linear scheme is that described in
\cite{rH99}. For the second modelling scheme a clear minimum occurs at
$4$. The local constant modelling scheme employs only lags that provide
an improvement to model prediction error and its error as a function of embedding window is
therefore monotonic (plateau onset occurs at $34$).  
For small values of embedding window the linear
scheme performs best, but for large values, behaviour is poor and
extremely erratic. Computation of description length utilising the
local constant scheme (solid line with circles) yields an optimal
embedding window of $15$. For clarity, the values $d_w=4,15,34$ are
marked as vertical dashed lines.}
\label{calceg}
\end{figure}

The example we consider is $2000$ points of the $x$ component of a
numerically integrated (sampling rate of $0.2$) trajectory of the
R\"ossler system, contaminated by additive Gaussian noise with a
standard deviation of $5\%$ of the standard deviation of the data. The
R\"ossler equations are given by
\begin{eqnarray}
\label{ross}
\left(\begin{array}{c}\dot{x}\\\dot{y}\\\dot{z}\end{array}\right)
& = & 
\left(\begin{array}{c}-y-z\\x+ay\\b+z(x-c)\end{array}\right)
\end{eqnarray}
where $a=0.398$, $b=2$ and $c=4$.  For these parameter values the data
exhibits broad band chaos. Figure \ref{calceg} demonstrates the
computation of (\ref{dl4}) as a function of embedding window. To
estimate model prediction error we employ the rather simple
interpolative scheme described in the previous section. For comparison, the performance of alternative
(more complex) modelling schemes is also shown in figure
\ref{calceg}. We find that alternative, more parametric, modelling
methods produce results which are sensitively dependent on ``correct''
choice of modelling algorithm parameters\footnote{By modelling
algorithm parameters we mean parameters associated with the model
selection scheme itself rather than only the parameters optimised by
that scheme.}.

The first zero of the autocorrelation function occurs at a lag of $8$
and the data exhibits a pseudo period of about $31$ samples. With the
embedding lag set at $8$, false nearest neighbours indicates a minimum
embedding dimension of $4$. Standard methods, therefore, suggest an
embedding window of roughly $32$.

By coincidence\footnote{In other examples, and for other amounts of
noise or with other lengths of data this proved not to be the case.},
the minimum of the model prediction error for a constant model occurs at
this value. Conversely, the minimum of the error of the local linear
model occurs at a value of $4$. This comparatively low value of
embedding window is due to the relative complexity of the local linear
modelling scheme \cite{cyclsurr}. Although this scheme performs best for
small embedding windows, the additional information introduced with
larger embedding windows is not recognised by this scheme. The main
reason for this is that the parameters of the scheme (neighbourhood
size, neighbourhood weights and so on) are also dependent on the
embedding dimension and embedding lag. For example, values of
neighbourhood size which work well for a small dimension embedding may
not work well for larger embedding dimension. Moreover, as embedding
dimension becomes larger it becomes difficult to find good values for
these parameters.. This general behaviour is observed in every example
we consider. Therefore, although the local linear scheme often provides
a good estimate of the optimal embedding {\em dimension} (as would false
nearest neighbours), the description length estimated from a local
constant model provides a much better estimate of the optimal embedding
{\em window}.

We have already mentioned that the local constant modelling scheme
selects only lags that provide some improvement in model prediction
error. Clearly, as $d_w$ increases there is a combinatorial
explosion. To address this combinatorial explosion is both difficult and
beyond the requirements of this algorithm. We consider only whether the addition of {\em successive}
lags offers an improvement. Suppose for a $d_e$ dimensional embedding the
 chosen model includes the lags $\{\ell_1,\ell_2,\ldots,\ell_k\}$ (where
$0\leq\ell_1\leq\ell_i<\ell_{i+1}\leq\ell_k<d_e$). To determine the
 set of model lags for the $(d_e+1)$-dimensional embedding we consider the
performance of the local constant model with lags
$\{\ell_1,\ell_2,\ldots,\ell_k,d_e\}$. If this model performs better
than the model with lags $\{\ell_1,\ell_2,\ldots,\ell_k\}$ then it is
accepted, otherwise we retain only the lags $\{\ell_1,\ell_2,\ldots,\ell_k\}$.

Therefore, the selected lags may be used as an
estimate of the optimal lags for a generalised variable embedding
(\ref{varemb}). In the case of the R\"ossler system data analysed in
figure \ref{calceg}, the optimal lags were $1$ to $15$ and $19$, $20$,
$24$, $26$, $29$, $32$ and $34$. Altogether, $22$ different
lags. Clearly, a $22$ dimensional embedding is excessive, and some subset
of these lags would probably prove sufficient. Moreover, the minimum
description length optimal embedding window is $15$, limiting the
selection to the first $15$ lags. It is reasonable to suppose that each
of these large number of lags may contribute some significant novel
information to the modelling scheme. However, the expression we hope to
optimise (\ref{dl4}) is independent of which lags are included (indeed,
in this example, they are {\em all} included\footnote{This is not the
case in general.}), and therefore we do not consider this issue more
closely here. We defer the selection of optimal lags from this set for
the modelling phase of dynamic reconstruction.

\begin{figure}[t]
\[\epsfxsize 130mm \epsfbox{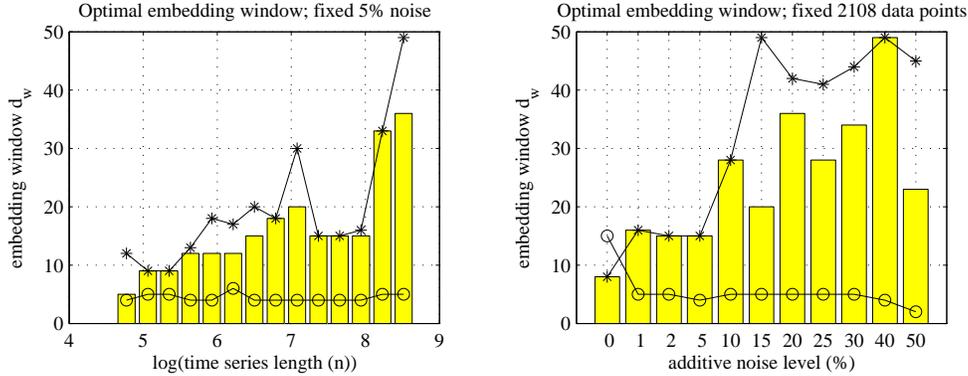}\]
\caption{{\bf Optimal embedding dimension as a function of data length
and noise level} The solid bars depict the optimal model size utilising
the methods described in this paper for a single realisation of R\"ossler time series data. The
panel on the left is for a fixed noise level of $5\%$ and time series
length between $118$ and $5000$ data. The panel on the right is for
fixed data length of $2108$ data and various noise levels (expressed as
percentage of the standard deviation of the data). For the cases where
noise was added to the time series, the results depicted here are for a
single realisation of that noise (not an average). This is the likely
cause of the moderate variation in the results observed for larger
noise levels. For comparison, the
embedding window that yielded minimum error for the local constant
(asterisks) and local linear (circles) models is also shown.}
\label{rossler}
\end{figure}

In figure \ref{rossler} we examine the effect of various noise levels
and different length time series on the selection of embedding
window. We observe that for longer time series, the optimal embedding
window is larger. This is consistent with what one might expect. For
short time series the optimal model can only capture the short term
dynamics and therefore only recent past history (a small embedding window)
is required. For larger quantities of data one is able to characterise
the more sensitive long term dynamics and a larger embedding window
provides significant advantage. Initially, an embedding window of about
$10$ is sufficient, while for the longest time series an embedding
window of $35$ is optimal. Significantly, these two values correspond to
approximately the first zero of the autocorrelation function (or
one-quarter of the pseudo-period) and the pseudo-period of the observed
time series. 

We note in passing, that, the optimal embedding window
for the local constant window is an upper bound on the minimum
description length best window. This is as we would expect. The
description length is the sum of a term proportional to the model
prediction error and a function which increase monotonically with
embedding dimension (the description length of the local constant
model). Therefore the minimum of the model prediction error must be no
less than the minimum of the description length.

Conversely, we find that the optimal embedding window
for the local linear method remains about $4$ or $5$ (roughly
corresponding to the optimal embedding dimension). 

\begin{figure}[t]
\[\epsfxsize 130mm \epsfbox{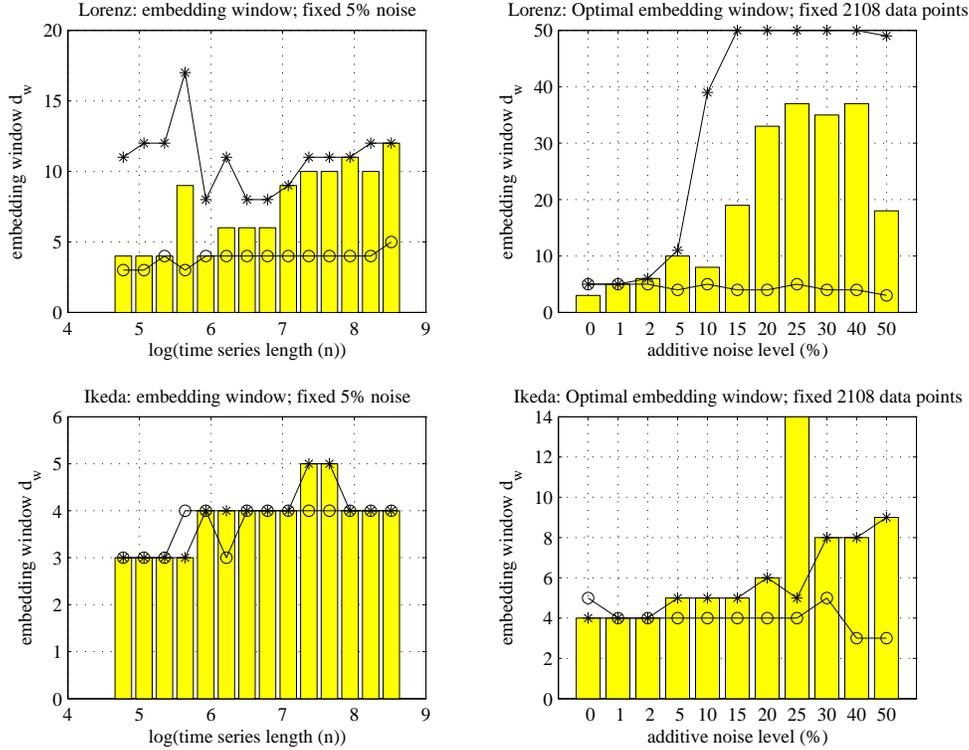}\]
\caption{{\bf Optimal embedding windows for Lorenz and Ikeda time
series.} The calculations depicted in figure
\ref{rossler} are repeated for time series of two standard systems. The top
two panels are for a single realisation of the chaotic Lorenz system,
the bottom two panels are for  a single realisation of
the chaotic Ikeda map. The solid bars depict the optimal model size
utilising the methods described in this paper. The leftmost panels are
for a fixed noise level of $5\%$ and time series length between $118$
and $5000$ data. The panels on the right are for fixed data length of
$2108$ data and various noise levels (expressed as percentage of the
standard deviation of the data). This is the likely
cause of the moderate variation in the results observed for larger
noise levels. For comparison, the embedding window
that yielded minimum error for the local constant (asterisks) and local
linear (circles) models is also shown.}
\label{ikedalorenz}
\end{figure}

Variation in the noise level for a fixed length time series demonstrates
similar behaviour. For noisier time series a larger embedding window is
required, as increasing the noise on each observation decreases the
useful information provided. As the information provided to the optimal
model by each observation decreases, more observations (a larger embedding
window) is required to provide all the available information. For noise
levels of up to $30\%$ this method provides consistent, repeatable,
results. Noisier time series tend to yield a larger variation in the
optimal estimates of embedding window. Note that in contrast, the
local linear scheme performs progressively worse, utilising a
diminishing window as the noise level is increased. We believe that this
is due to the additional parametric complexity of this modelling
method. As more noise is added to the data, the (relatively) complex
rules used to determine near neighbours and derive a weighted linear
prediction from these, becomes more prone to the system noise, and
actually performs worse.

In figure \ref{ikedalorenz} we repeat the above calculations for time
series generated from the standard chaotic Lorenz system and the Ikeda
map \cite{dK95}. Variation of optimal embedding window as a function of
noise and data length for the Lorenz data is very similar to the results
depicted in figure \ref{rossler} for the R\"ossler system. Increasing
noise level or time series length yields a larger optimal
model. Furthermore, optimal embedding window values tend to coincide
with the pseudo-period of the time series, or one-quarter, or one-half
of this value.

Results for the Ikeda map are substantially different. In this case the
optimal embedding window estimated  coincides with the value that
minimises the error of the local constant and linear models. In general,
an embedding dimension of $3$ or $4$ is suggested, and this is what one
would expect for this system\footnote{Although the fractal dimension of
the Ikeda map is less than two, a delay reconstruction of this map is
highly ``twisted'' and requires an embedding dimension of $3$ or $4$
to successfully remove all intersecting trajectories.}.

\begin{table}
\begin{center}
\begin{tabular}{r|c|c|c}
model & MDL & RMS & size\\\hline
Standard ($d_e=4$, $\tau=8$) & $-655\pm23$& $0.158\pm0.003$  & $15.6\pm2.9$\\
Windowed ($d_w=15$) & $-716\pm17$ & $0.151\pm0.004$  & $21.1\pm5.5$ \\
\end{tabular}
\end{center}
\caption{Comparison of model performance with standard constant lag
embedding (a {\em Standard Embedding}) and embedding over the embedding window suggested in figure
\ref{rossler} (a {\em Windowed Embedding}). Figures quoted are the mean of $60$ nonlinear models,
fitted with a stochastic optimisation routine to the same data set, and
standard deviations. Figures quoted here are for $2000$ data points with
$5\%$ noise, other values of these parameters gave similar, consistent,
results.
The three indicators are minimum description length
(MDL) of the optimal model, root-mean-square model prediction error
(RMS) and the model size (number of nonlinear terms in the optimal
model). For each indicator, the new embedding strategy shows clear
improvement. MDL and RMS have decreased, indicating a more compact
description of the data and a smaller prediction error,
respectively. Conversely, the mean model size has increased indicating
that more structure is extracted from the data. Several other measures
were also considered: mean amplitude of oscillation, correlation
dimension, entropy and estimated noise level \cite{effgka}. However, for
each of these measures the variance between simulations of models built
using the same embedding strategy was as large as that between the
different embedding strategies. The results of these calculations are therefore omitted.}
\label{modeltable}
\end{table}

We now return to the main purpose of estimating the embedding window,
namely the reconstruction of the dynamics. For the R\"ossler system
analysed in figures \ref{calceg} and \ref{rossler} we build nonlinear
models following the methods described in \cite{kJ98} with embedding
suggest by either autocorrelation and false nearest neighbours (namely
$d_e=4$ and $\tau=8$), hereafter referred to as a {\em Standard
Embedding}, or with the embedding window (of $34$), hereafter a {\em
Windowed Embedding}. Table
\ref{modeltable} compares the average model size (number of nonlinear
basis functions in the optimal model) and model prediction error for
$60$ models of this time series ($2000$ observations and $5\%$ noise)
with each of these two embedding strategies. These models are built to
minimise the description length of the data given the model, and therefore a
comparison of the optimal model description length is also given. These
qualitative measures show a consistent improvement in the model
performance for the model built from the windowed embedding.

\section{Applications}
\label{applications}

\begin{figure}[t]
\[\epsfxsize 125mm \epsfbox{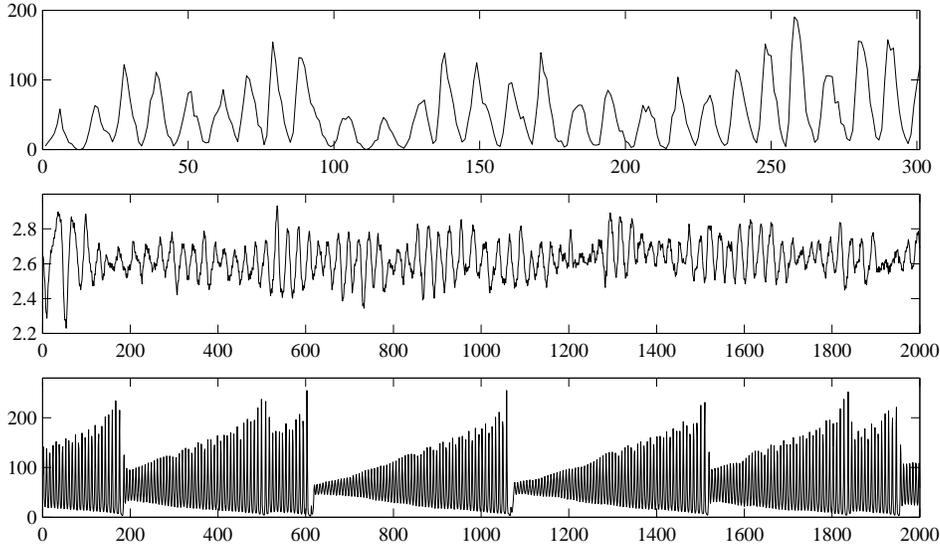}\]
\caption{{\bf Experimental time series data} The three experimental time
series examined in this paper are depicted (from top to bottom): annual
sunspot numbers for the period 1700 to 2000, a recording of human
electrocardiogram rhythm during ventricular fibrillation, and the
chaotic ``Santa-Fe'' laser times series. For the lower two panels only
the first 2000 points are utilised for time series modelling.}
\label{expdata}
\end{figure}

We now consider the application of this method to three experimental
time series: the annual sunspots times series \cite{hT90}, human electrocardiogram
(ECG) recordings of ventricular
fibrillation (VF) \cite{cic4,csf}, and experimental laser intensity data \cite{eW93,tS93}. The raw time series data
are depicted in figure \ref{expdata}. 

Since the main motivation for
selection of embedding window with the method described in this paper is
to improve modelling results we concentrate exclusively on the
comparison of the performance of nonlinear models of this data with
standard embedding techniques and the windowed embedding suggested by the
algorithm proposed here. By construction, the local constant modelling scheme
performs best with the windowed embedding. Therefore, we consider
a more complicated nonlinear radial basis modelling algorithm, first
proposed in \cite{kJ95a} and most recently described in
\cite{mdlnn}. Like the windowed embedding strategy, this modelling scheme is
designed to optimise the description length of the time series \cite{mdlnn}.

We are interested in two types of measures of performance: short term
behaviour (for example mean square prediction error) and dynamic behaviour
(invariant measures of the dynamical systems). Results equivalent to
those depicted in table \ref{modeltable} have also been computed and
are summarised in table \ref{exptable}.

\begin{table}
\begin{center}
\begin{tabular}{l|c|c|c|c}
data & MDL &  RMS & size &  CD\\\hline
sunspots & & & &\\
($d_e=6$, $\tau=3$) & $1267.9\pm12.1$ & $13.16\pm1.116$ &$7.32\pm1.818$ &
 $0.938\pm0.456$\\
 ($d_w=6$) & $1230.1\pm11.6$ & $12.31\pm0.6886$ & $6.96\pm1.50$ 
& $0.7836\pm0.4145$\\\hline
VF ECG  & & & &\\
($d_e=5$, $\tau=8$)& $-14333\pm28$ & $0.02468\pm0.0003846$  &
$31.47\pm7.326$ & $1.003\pm0.099$\\
 ($d_w=2$) & $-14163\pm14$ & $0.02751\pm0.0001500$ & 
$15.38\pm2.66$ & $1.124\pm0.107$\\\hline
laser  & & & &\\
($d_e=5$, $\tau=2$) & $5753.6\pm153.9$ &
$2.405\pm0.2954$ &  $100.8\pm12.3$ & n/a\\
 ($d_w=10$) & $5239.8\pm159.0$ & $1.767\pm0.1992$ & $109.5\pm12.3$ 
& $0.8637\pm0.7999$
\end{tabular}
\end{center}
\caption{Comparison of model performance with standard constant lag
embedding and embedding over the embedding window suggested in figure
\ref{rossler}. Figures quoted are the mean of $60$ nonlinear models,
fitted with a stochastic optimisation routine to the same data set, and
standard deviations. Figures quoted here are for $2000$ data points,
where more data is available, longer time series samples gave  similar, consistent,
results.
The four indicators are minimum description length
(MDL) of the optimal model, root-mean-square model prediction error
(RMS), the model size (number of nonlinear terms in the optimal
model), and the correlation dimension (CD) of the free run dynamics. For
the laser time series, none of the models built using the standard
embedding produced stable dynamics and
it was therefore not possible to estimate correlation dimension. The
correlation dimension estimated directly from these three data sets was
$0.396$, $1.090$, $1.182$ (note that the low value for the first data
set is an artifact of the short time series).}
\label{exptable}
\end{table}

Table \ref{exptable} shows that for the sunspot time series and the
experimental laser intensity recording, the windowed embedding improved model performance. That is, the
description length was lower, the one-step model prediction error was
less and the models were larger. However, with the exception of one step
model prediction error the difference in these measures was not
statistically significant. For the recording of human VF the new method
did not improve model performance and, in fact, the optimal embedding
window was $d_w=2$: substantially smaller  than one would reasonably
expect from such a complex biological system. It seems plausible, that in
this case, the time series under consideration is too short, noisy or
non-stationary (this conclusion is supported by figure
\ref{expdata}). Finally, we note that the result for the sunspot time
series is particularly encouraging because this improvement in short term
predictability is achieved with a much smaller embedding ($d_w=6$
compared to $d_e\tau=18$).

However, as has been observed elsewhere \cite{mdlnn}, short term
predictability is not the best criteria with which to compare models of
nonlinear dynamical systems. Therefore, for each model we estimated
correlation dimension, noise level and entropy, using a method described
in \cite{effgka}. Furthermore, under the premise that these models should
exhibit pseudo-periodic dynamics we also computed mean limit cycle
diameter (i.e. the amplitude of the limit cycle oscillations. In every
case we found that the dynamics exhibited by models built from the
traditional (i.e. uniform) embedding strategy was more likely to either
be a stable fixed point or divergent.

\begin{figure}[pt]
\[\epsfxsize 125mm \epsfbox{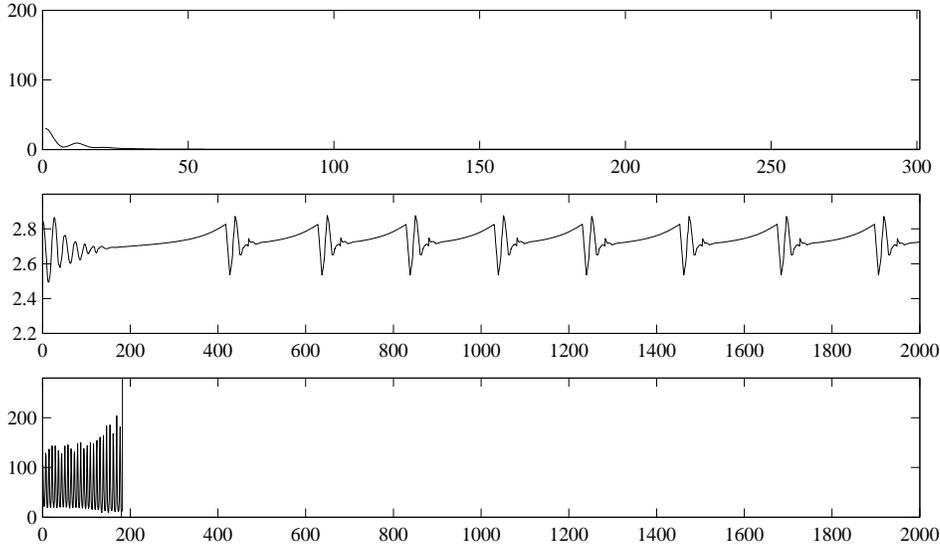}\]
\caption{{\bf Typical model behaviour using the standard embedding strategy $(d_e,\tau)$:} Three simulated time
series from models of the experimental data examined in this paper are
depicted. The panels correspond to those of figure \ref{expdata} and the horizontal and vertical axes in these
figures are fixed to be the same values as the corresponding panels of
figure \ref{expdata}. }
\label{badruns}
\end{figure}

\begin{figure}[pt]
\[\epsfxsize 125mm \epsfbox{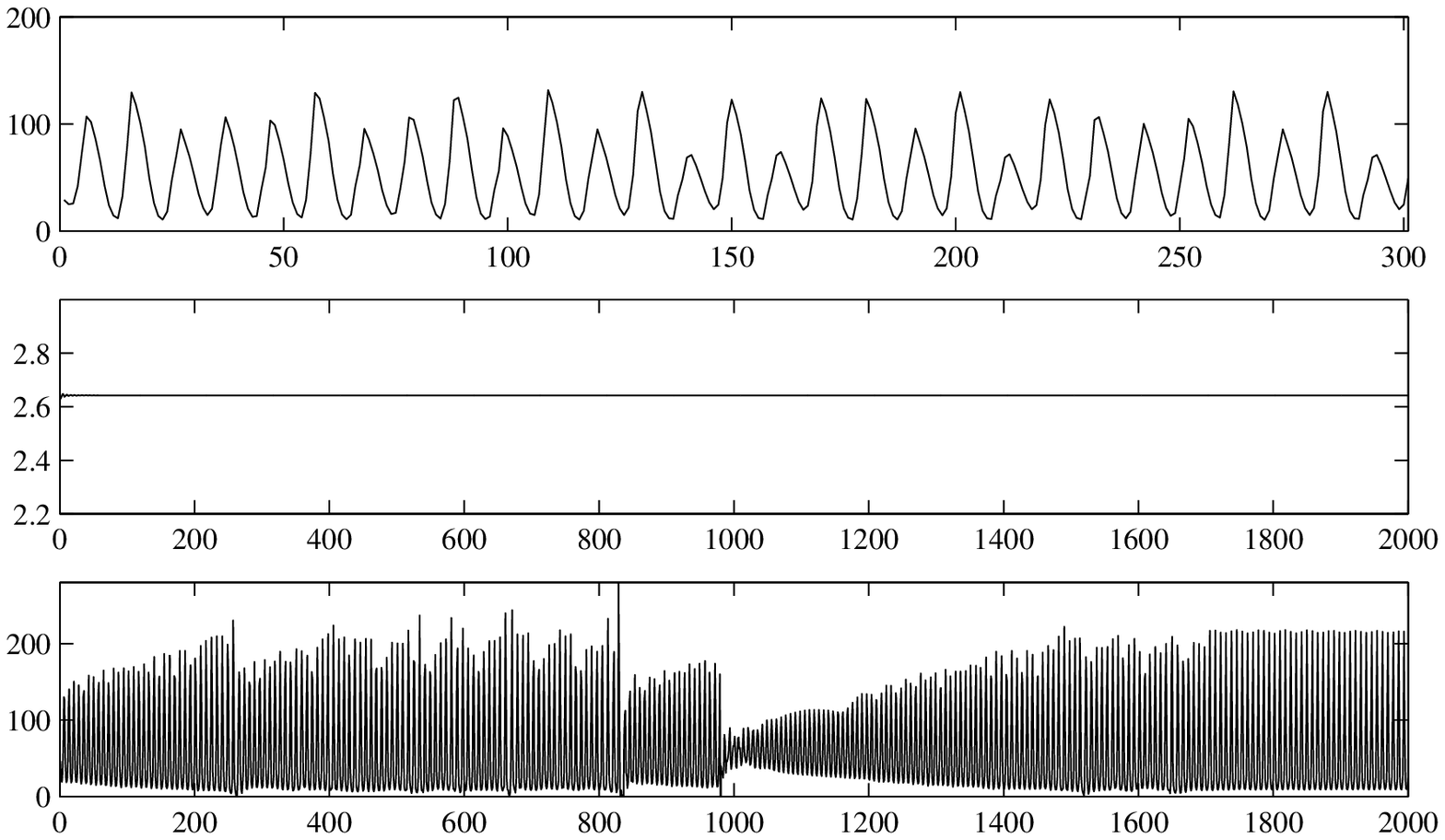}\]
\caption{{\bf Typical model behaviour using the windowed embedding strategy $(d_w)$:} Three simulated time
series from the experimental data examined in this paper are
depicted. The panels correspond to those of figure \ref{expdata} and the horizontal and vertical axes in these
figures are fixed to be the same values as the corresponding panels of
figure \ref{expdata}. }
\label{goodruns}
\end{figure}

\begin{figure}[pt]
\[\epsfxsize 125mm \epsfbox{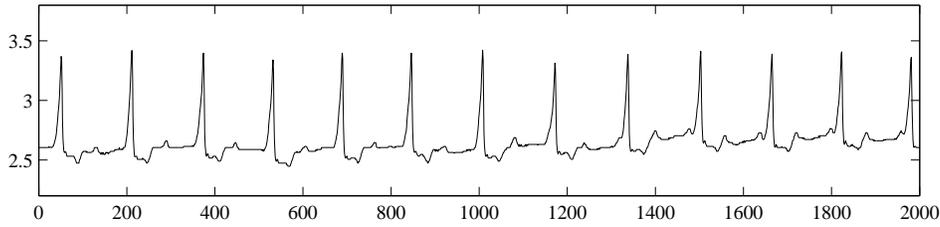}\]
\caption{{\bf Observed electrocardiogram trace after resuscitation:}
Following the episode of VF depicted in figure \ref{expdata}, electrical
resuscitation was successfully applied. The waveform depicted in this
figure was observed 120 seconds after application of
defibrillation. Notice that the waveform observed here is similar to the
asymptotic behaviour of the model depicted in figure \ref{badruns}. Both
the period and the shape of these time series are similar.}
\label{recover}
\end{figure}

Finally, figures \ref{badruns} and \ref{goodruns} show typical noise free
dynamics in models of each of these three systems. No effort was made to
ensure that the models performed well and the models and simulations
presented in these figures were selected at random. For the sunspots and
laser dynamics  (top and bottom panels) the new method clearly performs
better. Typically, the original method produced laser dynamics and
sunspots simulations that were divergent and a stable fixed point
(respectively). These results are typical. In contrast the windowed method
yields models which exhibit bounded (almost) aperiodic
dynamics\footnote{Closer examination of the laser dynamics indicates that
it eventually settles to a stable periodic orbit (this phenomenon can be
observed toward the end of the time series depicted in figure \ref{goodruns}.}. 

Using the windowed embedding we found that the long term dynamics for
models of the VF data performed badly. However, this is to be expected as
the optimal embedding window was $2$. For this data none of the models
produced with either method performed well. Hence, with this short,
noisy and non-stationary data the models (built with either embedding
strategy) failed to capture the underlying dynamics. Significantly, the
estimate of $d_w$ provided by our algorithm indicated that this would be
the case. We found that the optimal embedding is only two dimensional and
this suggests that the best thing to do is to not build a model of the
dynamics at all. Hence, even this negative result is encouraging: we
have established the limitations of this algorithm and found that even
in this situation the results are consistent. We note with some
curiosity that the dynamics exhibited in the second panel of figure
\ref{badruns} closely resembles the human electrocardiogram during
regular rhythm (see figure \ref{recover} for a representative recording), despite the data being recorded during VF --- we have no
explanation for this. This behaviour, although exhibited by this one
model was not present in simulations from all similar models. It is
suggestive that the VF waveform includes information concerning the
underlying (slower) sinus rhythm dynamics. But the evidence for this is
definitely not conclusive.

\section{Conclusions}
\label{conclusions}

We have approached the problem of optimal embedding from a modelling
perspective. In contrast to previous studies (which focused on
estimating dynamic invariants) our primary concern was
selection of embedding parameters that provide the optimal
reconstruction of the underlying dynamics for an observed time series.
To achieve this we assumed that the optimal model is that which
minimises the description length the data. From this foundation we
showed that the best embedding has a constant lag ($\tau=1$) and a
relatively large embedding window $d_w$. In general the optimal $d_w$
will be determined by the amount of noise and the length of the the time
series. From an information theoretic perspective this is what one would
expect: $\tau>1$ implies some information is missing from the
embedding. The optimal value of $d_w$ reflects a balance between a small
embedding with two little information to reconstruct the dynamics and a
large embedding where the model ceases to describe the dynamics.

To compute the quantity $d_w$ we introduced an extremely simple {\em
non-predictive} local constant model of the data and select the value of
$d_w$ for which this model performs best. One can see that this offers a
new and intuitive method for selection of embedding
parameters. In essence, one could neglect description length and simply
choose the embedding such that this model performs best. However, the
addition of description length makes the optimal $d_w$ dependent not
only on the noise but also on the length of the time
series. We see that for short time series one shouldn't be confident of
a large embedding window.

The similarity between this new method of embedding window selection and
the well established false nearest neighbour technique \cite{lC97} is
more than superficial\footnote{The comparison of this method to that
described in \cite{lC97} is particularly apt. Cao introduces a modified
false nearest neighbour approach which, like our method, avoids many of
the subjective parameters of alternative techniques.}. In section \ref{examples}
and \ref{applications} we provided an explicit comparison to between our technique
and the ``standard'' false nearest neighbour method. However, there are
various improvements to this algorithm (such as \cite{lC97}) which are
worthy of further consideration. Nonetheless, there are
several important distinctions between our method, and these
false nearest neighbour techniques. As we have already emphasised, the aim of this
method (to achieve the best model of the dynamics) differs from that of
false nearest neighbours (topological unfolding). Furthermore, the
incorporation of minimum description length means that our method
explicitly penalises for short or noisy time series. 

At a functional level, the two algorithms are similar because both
methods seek to avoid data points which are close, but which quickly
diverge. Such points are (respectively) either false nearest neighbour
of bad nonlinear predictors of one another. However, where as false
nearest neighbour methods seek only to avoid this situation
(i.e. spreading out the data is sufficient), the windowed embedding
method insists that the neighbours which are the best predictors
be found. 

Consider the situation where a systems' dynamics are either
stochastic or extremely high dimensional. Using false nearest neighbour
methods, one may simply embed the data in a high enough dimension so
that the data are sufficiently sparse. However, doing so does not
improve the nonlinear prediction error, consequently, the windowed embedding
method would prefer a small embedding window.

Conversely, consider the situation at a seperatrix. Points which are
close do rapidly diverge from one another and so they will appear as
false near neighbours for large embedding dimension, until (at a time
scale similar to that of the underlying system) the points are
eventually, sufficiently spread). But from a nonlinear prediction 
view-point, these points are equally difficult to predict for all
embedding dimension, and again the windowed embedding method will
indicate a much smaller embedding dimension than that suggested by a
strict application of false nearest neighbours\footnote{We acknowledge
that this problem is actually related to the ``plateau'' observed in
plots of the fraction of false nearest neighbours against embedding
dimension. In many cases, prudent selection of ``plateau-onset'' can
minimise the problem. However, this remains somewhat subjective.}.
 
Finally, we note that the examples of section \ref{examples} showed that this method
performed consistently and the applications in section
\ref{applications} showed that selecting embedding parameters in this
way improved the model one-step prediction error. In effect, this is a
demonstration that the method is working as expected. More
significantly, we found that the dynamics produced by models built from
windowed embedding also behaved more like the experimental dynamics than
for models built from a standard embedding. This is a very positive
results, however, we are now faced with a more substantial problem. The
problem of building the best nonlinear model for the data once the
embedding window has been determined \cite{mdlnn}. Information theory
has shown us that the optimal embedding should fix $\tau=1$, we now need
to consider the practice of nonlinear modelling to determine which lags
$\ell=1,2,3,\ldots,d_w$ are significant for practical reconstruction
from specific experimental systems.

\section*{Acknowledgments}

This work was supported by a Hong Kong Polytechnic University Research
Grant (No. A-PE46).

\bibliographystyle{unsrt}

\end{document}